\documentclass[twocolumn,aps,pra,groupedaddress,superscriptaddress,amsmath,amssymb,noeprint,nofootinbib,floatfix,10pt]{revtex4-2}
\usepackage[T1]{fontenc}
\usepackage{graphicx}
\graphicspath{{./Figures/}}
\usepackage{dcolumn}
\usepackage{bm}
\usepackage[dvipsnames,table,xcdraw]{xcolor}
\usepackage{hyperref}
\hypersetup{colorlinks,linkcolor={MidnightBlue},citecolor={MidnightBlue},urlcolor={MidnightBlue}}  
\usepackage{url} 
\usepackage{tabularray}

\usepackage{wrapfig}

\usepackage{tensor}

\usepackage{physics}
\usepackage{amsfonts}
\usepackage{verbatim}
\usepackage{amsmath}
\usepackage{csquotes}
\usepackage{color}
\usepackage{soul}
\usepackage{amsthm}
\usepackage{bm}
\usepackage{graphicx}
\usepackage[normalem]{ulem}
\usepackage{mathtools}

\usepackage{yfonts}

\usepackage{verbatim}

\newenvironment{equations}
{\begin{equation}\begin{aligned}}
		{\end{aligned}\end{equation}\ignorespacesafterend}

\newenvironment{equations*}
{\begin{equation*}\begin{aligned}}
		{\end{aligned}\end{equation*}\ignorespacesafterend}

\newcommand{\prt}[1]{\left(#1\right)}

\newcommand{\CP}{\textrm{CP}}

\newcommand{\intmp}{\int_{-\infty}^{+\infty}}

\newcommand{\Schr}{Schr\"{o}dinger\ }

\newcommand{\dg}{\dagger}
\newcommand{\tl}{\tilde}

\newcommand{\prtq}[1]{\left[#1\right]}

\newcommand{\prtg}[1]{\left\{#1\right\}}

\newcommand{\prtB}[1]{\Bigg(#1\Bigg)}

\newcommand{\tmcA}{\tilde{\mathcal{A}}}

\newcommand{\tmcL}{\tilde{\mathcal{L}}}

\newcommand{\mcA}{\mathcal{A}}

\newcommand{\mcC}{\mathcal{C}}

\newcommand{\mcH}{\mathcal{H}}

\newcommand{\mcM}{\mathcal{M}}

\newcommand{\mcT}{\mathcal{T}}
\newcommand{\mcU}{\mathcal{U}}

\newcommand{\mbE}{\mathbb{E}}

\newcommand{\mbP}{\mathbb{P}}

\newcommand{\hA}{\hat{A}}

\newcommand{\hH}{\hat{H}}

\newcommand{\hK}{\hat{K}}
\newcommand{\hL}{\hat{L}}
\newcommand{\hM}{\hat{M}}

\newcommand{\hO}{\hat{O}}

\newcommand{\hU}{\hat{U}}

\newcommand{\hW}{\hat{W}}
\newcommand{\hX}{\hat{X}}

\newcommand{\hZ}{\hat{Z}}

\newcommand{\hmu}{\hat{\mu}}

\newcommand{\tL}{\tilde{L}}

\newcommand{\bx}{\mathbf{x}}
\newcommand{\by}{\mathbf{y}}

\newcommand{\br}{\mathbf{r}}

\makeatletter 

\renewcommand\onecolumngrid{
\do@columngrid{one}{\@ne}%
\def\set@footnotewidth{\onecolumngrid}
\def\footnoterule{\kern-6pt\hrule width 1.5in\kern6pt}%
}

\renewcommand\twocolumngrid{
\def\footnoterule{
\dimen@\skip\footins\divide\dimen@\thr@@
\kern-\dimen@\hrule width.5in\kern\dimen@}
\do@columngrid{mlt}{\tw@}
}%

\makeatother    

\begin{document}

\title{Non-Markovian Poissonian Spontaneous Collapse Models}

\author{Nicol\`{o} Piccione}
\email{nicolo.piccione@units.it}
\affiliation{Department of Physics, University of Trieste, Strada Costiera 11, 34151 Trieste, Italy}
\affiliation{Istituto Nazionale di Fisica Nucleare, Trieste Section, Via Valerio 2, 34127 Trieste, Italy}



\begin{abstract}
Spontaneous collapse models provide a possible solution to the measurement problem by modifying standard quantum dynamics. 
The modification consists of adding non-linear and stochastic terms inducing wavefunction collapse in space.
Non-Markovian versions of these models are motivated by physical reasons, phenomenological consistency, and potential for relativistic extensions.
Here, we investigate a non-Markovian version of the Poissonian Spontaneous Localization (PSL) model, i.e., a model characterized by instantaneous and localized collapse events.
We assume that our model is characterized by a typical time scale $\tau_C$ so that, given an initial state $\rho_0$ of standard quantum matter at time $t=0$, we derive an effective long-time ($t\gg \tau_C$) statistical dynamics in terms of a CPTP map $\Phi_t$.
We then show how $\Phi_t$ can be made equal to that obtained by non-Markovian CSL models.
Moreover, given $\Phi_t$, we obtain the associated time-convolutionless master equation by means of a supercumulant expansion.
Finally, we characterize the collapse events process (for events with $t \gg \tau_C$) given an initial quantum state $\rho_0$ at $t=0$.
\end{abstract}

\maketitle

\section{Introduction\label{Sec:Introduction}}

Quantum mechanics is extraordinarily successful, yet its standard formulation is afflicted by the so-called measurement problem: the linear and unitary evolution prescribed by the Schrödinger equation does not, by itself, explain the emergence of definite outcomes in individual experiments~\cite{Book_Bell2004Speakable,Book_Norsen2017foundations,Book_Durr2020understanding,Book_Tumulka2022Foundations}. Spontaneous collapse models address this issue by modifying the standard quantum dynamics: non-linear, stochastic terms are postulated and added to the Schrödinger dynamics. These terms, negligible for microscopic systems but dominant for macroscopic ones, suppress macroscopic superpositions while reproducing ordinary quantum predictions at the microscopic level~\cite{Review_Bassi2003Dynamical,Review_Bassi2013Models,Book_Tumulka2022Foundations,Review_Bassi2023CollapseModels}.

The most studied collapse models are Markovian~\cite{Ghirardi1986Unified,Diosi1987Universal,Diosi1989Models,Pearle1989CSL,Ghirardi1990_CSL,Review_Bassi2003Dynamical,Review_Bassi2023CollapseModels}, mainly because of their simplicity compared to their non-Markovian counterparts~\cite{Adler2007NonWhiteNoises,Adler2008NonWhiteNoisesII}. In fact, while the Markovian assumption is mathematically convenient, there are strong reasons to investigate non-Markovian extensions. For example, assuming a white-in-time noise that affects charged particles leads to a divergent radiation power~\cite{Donadi2014RadiationEmission,Donadi2015Radiation,Donadi2021NovelCSLBounds,Donadi2021UndergroundTest}, while the same does not hold for the non-Markovian versions of the same models~\cite{Donadi2014RadiationEmission,Donadi2015Radiation}.\footnote{In particular, for Markovian collapse models~\cite{Piccione2025ExploringMassDependence}, the per-frequency radiation emission rate is proportional to $1/\omega$, where $\omega$ is the frequency of the emitted light, and we recall that $\int_0^{\infty} \dd{\omega} \omega^{-1} = \infty$.}
Moreover, several works have indicated that non-Markovianity is necessary for any serious attempt to construct a relativistic collapse model~\cite{Myrvold2017RelativisticMarkovian,Jones2021MassCoupled,Diosi2022RelativisticGKSLNope,Gundhi2025RelativisticCollapse}. In fact, several attempts in this direction lead to a non-Markovian evolution of the quantum state~\cite{Tumulka2006relativistic,Tumulka2009rGRWf,Bedingham2011Relativistic}. For all these reasons, non-Markovian versions of continuous collapse models have been thoroughly investigated both from the theoretical~\cite{Adler2007NonWhiteNoises,Adler2008NonWhiteNoisesII,Bassi2009NonMarkovianTrajectories,Diosi2014NonMarkovianGaussianDynamics,Tilloy2021NonMarkovianWave} and phenomenological~\cite{Carlesso2018ColoredCSLExperimentalTests,Toros2017CdCSLBounds,Toros2018BoundsCalculations} perspectives.

The above motivations also apply to the Poissonian Spontaneous Localization (PSL) model~\cite{Piccione2023Collapse,Piccione2025ExploringMassDependence}. 
This model is built by first positing that spacetime is randomly filled with special points called Collapse Points (CPs). 
A qubit, initialized in $\ket{0}$, is associated with each CP, and $\ket{1}$ denotes the state orthogonal to $\ket{0}$.
Each CP-qubit interacts with standard quantum matter and is then subjected to a spontaneous measurement of $\dyad{1}$, leading to an induced localization of standard quantum matter when found in $\ket{1}$.
Following the standard terminology in the Foundations of Physics~\cite{Book_Tumulka2022Foundations}, we call such an occurrence a \emph{flash}.
The PSL model is obtained by studying the limit regime in which the density of CPs goes to infinity while their coupling with standard quantum matter goes to zero, in a somewhat formally similar way to the continuous limit of collision models~\cite{Review_Ciccarello2022CollisionModels}. In the Markovian version of PSL~\cite{Piccione2023Collapse,Piccione2025ExploringMassDependence}, the interaction time between a CP-qubit and standard quantum matter is assumed to be instantaneous, leading to dynamics similar to those of the spontaneous collapse model of Ref.~\cite{Tumulka2006spontaneous}, another collapse model with flashes.
Notably, the flashes of PSL can be used as sources of classical Newtonian gravity with a stable vacuum~\cite{Piccione2025NewtonianPSL,Piccione2026EnergyIncreaseGPSL}, providing a hybrid classical-quantum theory of Newtonian gravity in which the reduced dynamics of quantum matter is, nevertheless, Markovian.

In this work, we develop a non-Markovian extension of PSL. Our strategy is to retain the spacetime picture underlying the original model while allowing each CP to interact with quantum matter on a finite time scale $\tau_C$. Under certain assumptions, which should be valid when considering long-time dynamics ($\gg \tau_C$), we find the map $\Phi_t$ that describes the average reduced dynamics of quantum matter. We then derive the associated time-convolutionless master equation by means of the supercumulant expansion~\cite{Book_Breuer2002,Szankowski2023OpenQuantumSystems,Szankowski2025OpenQuantumSystems}, clarifying its connection with the exact map and with previously known non-Markovian Gaussian dynamics~\cite{Adler2007NonWhiteNoises,Diosi2014NonMarkovianGaussianDynamics}.
Finally, given an initial state $\rho_0$ at time $t=0$, we characterize the stochastic flash process when considering flash locations with $t\gg \tau_C$. That is, we provide the formula for the joint intensity density\footnote{Heuristically, the intensity $\mu_F^{(n)} (x_1,\dots,x_n|\rho_0)$ gives the probability $\mu_F^{(n)} (x_1,\dots,x_n|\rho_0)\dd[4]{x_1}\cdots\dd[4]{x_n}$ of flashes occurring in the infinitesimal neighborhood of the points $x_1,\dots,x_n$, regardless of the presence of other flashes elsewhere in spacetime. While the quantities $\mu_F^{(n)} (x_1,\dots,x_n|\rho_0)$ are not probability densities, they completely characterize the stochastic flash process (see Appendix~\ref{APPSec:PointProcessMathematics}).} $\mu_F^{(n)} (x_1,\dots,x_n|\rho_0)$ of the occurrence of $n$ flashes at points $x_1,\dots,x_n$ where $t_j \gg \tau_C$ for $1\leq j \leq n$.

\section{Building the model\label{Sec:ModelBuilding}}

The goal is to build a non-relativistic, non-Markovian extension of the spontaneous collapse model presented in Ref.~\cite{Piccione2023Collapse} and investigated in Refs.~\cite{Piccione2025ExploringMassDependence,Piccione2025NewtonianPSL,Piccione2026EnergyIncreaseGPSL}. 
We start by fixing some terminology and notation and presenting the ideas behind the model. In the following, we set $\hbar=1$. 

We assume that (Galilean) spacetime is randomly filled with special points that we call Collapse Points (CPs). These points are distributed according to a Poissonian distribution so that in every infinitesimal spacetime volume $\dd[4]{x}$ there is a probability $\mu_C \dd[4]{x}$ of finding one of these special points, where $\mu_C$ is the spacetime density of the CPs. In fact, a Poissonian sprinkling~\cite{Review_Surya2019CausalSetApproach} with constant density assures us that the distribution of such points is invariant for Galilean and Lorentz transformations~\cite{Review_Surya2019CausalSetApproach,Bedingham2021Collapse}.\footnote{Even more, in a general Lorentzian manifold, the Poisson sprinkling is coordinate-invariant and diffeomorphism-invariant, a fact which has also been used to propose prototypical versions of spontaneous collapse models in curved spacetimes~\cite{Bedingham2016RelativisticCollapseBlackHole,Juarez2018CollapseModelsHadamardCondition}.
We stress, however, that only the underlying sprinkling process has this relativistic invariance; the dynamics investigated in the paper is non-relativistic.}
To each of these CPs we associate an internal Hilbert space that we take to be that of a two-level system, i.e., a qubit. 
Initially, at $t=-\infty$, each of these qubits is initialized in the state $\ket{0}$ and interacts with quantum matter by means of an interaction Hamiltonian $\sqrt{\Upsilon}\hH_{x_C} (t)$, where $x_C := (t_C,\bx_C)$ denotes the spacetime location of the CP and $\sqrt{\Upsilon}$ quantifies the strength of the interaction.\footnote{The square-root has been introduced for later convenience.}
The interaction Hamiltonian is such that $\hH_{x_C} (t) = 0$ for $t>t_C$ and, at $t=t_C$, the CP-qubit undergoes a spontaneous (projective) measurement in the basis $\prtg{\ket{0},\ket{1}}$; if the result is $\ket{1}$, we say, following standard terminology in the Foundations of Physics~\cite{Book_Tumulka2022Foundations}, that a \emph{flash} occurred at spacetime location $x_C$.

The dynamics of the full model can be described as follows. Consider CPs at spacetime locations $x_1,x_2,\dots$ where $t_j>t_0$ for all CPs\footnote{For any given time $t_0$, the probability of having zero CPs with $t_C=t_0$ is one.} and an initial state $\sigma_0$ at time $t_0$. This is the state of both standard quantum matter and CPs in the future of $t_0$.
In fact, since CPs with $t_C < t_0$ have already collapsed, the effect of their spontaneous collapse is already encoded in $\sigma_0$ and they can be neglected.
Moreover, being projected either to the pure state $\ket{0}$ or $\ket{1}$, CP-qubits in the past of $\sigma_0$ cannot be entangled with quantum matter or other CP-qubits.
On the other hand, the initial state $\sigma_0$ is, in general, an entangled state of standard quantum matter and CP-qubits in the future of $t_0$. Between spontaneous measurements of the CP-qubits, the state evolves according to the total Hamiltonian
\begin{equation}\label{eq:TotalHamiltonian}
\hH (t) = \hH_S +\sqrt{\Upsilon} \sum_n \hH_{x_n} (t),
\end{equation}
where $\hH_S$ is the standard Hamiltonian of quantum matter, $x_n$ denotes the spacetime location of the $n$-th CP, and we assume the following structure:
\begin{equation}\label{eq:InteractionHamiltonianStructure}
\hH_{x_n} (t) = \hL_{x_n} (t) \otimes \hX_{n}.
\end{equation}
In the above equation, $\hX_{n}$ is the standard Pauli operator $\hX$ acting on the $n$-th CP-qubit,\footnote{One has that $\hX=\dyad{0}{1}+\dyad{1}{0}$.} and $\hL_{x_n} (t)$ is a time-dependent operator acting on the quantum matter Hilbert space. Moreover, we assume that there exists a time $\tau_C$ such that for $\abs{t-t_n}\gg \tau_C$ we have $\hL_{x_n} (t) \sim 0$.
The evolution between spontaneous CP-qubit measurements is therefore given by
\begin{equation}\label{eq:TotalUnitaryEvolution}
\hU (t',t) := \mcT_{\hH} \exp{-i\int_{t}^{t'} \hH (s)\dd{s}},
\end{equation}
where $\mcT_{\hH}$ denotes time-ordering of the operators $\hH (s)$.

An important physical regime is achieved when $\Upsilon \to 0$, $\mu_C \to \infty$, and $\Upsilon \mu_C \to \lambda >0$. Hereafter, we collectively denote this set of limits as the PSL limit. Obtaining meaningful equations in the PSL limit will be the main focus of this paper.

\section{Unitary unraveling of quantum matter dynamics\label{Sec:UnitaryUnravelingCPQM}}

The goal of this section is to provide a unitary unraveling of the dynamics of standard quantum matter in the long-time limit on the timescale set by $\tau_C$.
In the following, we set $t_0=0$ to lighten the notation.

Starting from Eq.~\eqref{eq:TotalHamiltonian} and moving to the interaction picture with respect to $\hH_S$, we get\footnote{We omit the symbol ``$\otimes$'' when there is no risk of confusion.}
\begin{equation}
\hH_I (t) = \sum_n \tL_{x_n} (t) \hX_n,
\quad
\tL_{x_n} (t) := \hU_S^\dg (t)\hL_{x_n} (t) \hU_S (t)
\end{equation}
where $\hU_S (t) :=\exp{-i t \hH_S}$. The unitary evolution associated with $\hH_I (t)$ is
\begin{equation}
\hU_I (t) = \mcT_{\hH_I} \exp{-i \sqrt{\Upsilon}\int_{0}^{t} \hH_I (s) \dd{s}},
\end{equation}
where $\mcT_{\hH_I}$ denotes time-ordering of the operators $\hH_I (s)$.
Before proceeding, let us introduce some convenient notation for dealing with super-operators. In general, we associate to each operator $\hA$ (capital letter with the hat) its super-operator counterpart $\mcA$ (the same capital letter in calligraphic font). 
Thus, we introduce the following super-operators\footnote{We are following the notation of Refs.~\cite{Szankowski2023OpenQuantumSystems,Szankowski2025OpenQuantumSystems}. The symbol ``$\bullet$'' is used as a placeholder for the
argument of superoperator when it acts on operators on its right. For example, $\mcH_I (t) := \comm{\hH (t)}{\bullet}$ means that $\mcH_I (t) \rho_t = \comm{\hH (t)}{\rho_t}$.} for the interaction-picture dynamics:
\begin{equation}
\mcH_I (t) := \comm{\hH_I (t)}{\bullet},
\qquad
\mcU_I (t) := \hU_I (t) \bullet \hU_I^\dg (t).
\end{equation}
One can check that\footnote{The adjoint of a super-operator is defined with respect to the Hilbert-Schmidt inner product~\cite[page 15]{Szankowski2025OpenQuantumSystems}.} $\mcU_I^\dg (t) = \mcU_I^{-1}(t)$. Defining $\mcH_{x_n} (t):= \comm{\hH_{x_n} (t)}{\bullet}$ we see that
\begin{equation}
\mcH_I (t) = \mcU_S^\dg (t)\prt{\sum_n \mcH_{x_n} (t)}\mcU_S (t),
\end{equation}
where $\mcU_S (t):= \hU_S (t) \bullet \hU_S^\dg (t)$.
We also have that
\begin{equation}
\mcU_I (t) = \mcT \exp{-i \sqrt{\Upsilon}\int_{0}^{t} \mcH_I (s) \dd{s}}.
\end{equation}
Hereafter, we will omit the subscript $I$ as all operators and super-operators will be considered to be in interaction picture if not otherwise stated.

Given the disposition of CPs and the initial state $\sigma_0$, a mathematically equivalent way to obtain the final state $\sigma_{t}$ at time $t>0$ is the following. Using Eq.~\eqref{eq:TotalUnitaryEvolution} with $\mcU (t) := \hU (t,0) \bullet \hU^\dg (t,0)$ gives a unitary evolution of both quantum matter and CP-qubits up to time $t$. Instead of considering the measurements of the CP-qubits with $t_C \in [0,t]$ to occur at their time locations, we can equivalently measure all of them (i.e., those with $0 < t_C \leq t$) at the final time $t$. This is possible because, for a CP located at $x_C$, the interaction vanishes for $t>t_C$.
Then, a possible final state is obtained by performing a projective measurement of $\hZ=\dyad{1}-\dyad{0}$ for all CP-qubits whose CPs are located between the $t=0$ and $t>0$ constant time hyperplanes in spacetime.
Let us denote by $Z_t$ the outcome of such a measurement, by $P(Z_t)$ the probability of its realization,\footnote{This probability, as well as other quantities in this section, can be made well-defined by first considering only CPs within a bounded region of spacetime, so that there is a finite number of CPs, and then taking the limit to infinite volume.} and by $\hat \Pi_{Z_t}$ the projection operator associated to the outcome $Z_t$. Then, the final state at time $t$ based on the spontaneous measurements results between $t=0$ and $t$ is
\begin{equations}
\sigma_t^{(Z_t, \CP)} &= \frac{1}{P(Z_t)}\hat \Pi_{Z_t} \prt{\mcU_\CP (t) \sigma_0} \hat \Pi_{Z_t},
\\
P(Z_t) &= \Tr{\hat \Pi_{Z_t} \prt{\mcU_\CP (t) \sigma_0} \hat \Pi_{Z_t}},
\end{equations}
where we inserted subscript $(Z_t,\CP)$ to remind us that the final state depends both on the outcomes of the spontaneous measurements as well as the spacetime locations of the CPs. Indeed, we also added the subscript $\CP$ to the unitary evolution, as it also depends on the location of the CPs.
Averaging over the possible measurement outcomes gives the final statistical state
\begin{equation}
\sigma_t^{(\CP)} 
= 
\mbE_{Z_t} \prtq{\sigma_t^{(Z_t, \CP)}}
=
\sum_{Z_t} \hat \Pi_{Z_t} \prt{\mcU_\CP (t) \sigma_0} \hat \Pi_{Z_t},
\end{equation}
where $\mbE_Y$ denotes expectation over the stochastic process $Y$. It follows that the reduced state of quantum matter at time $t$, for a given realization of the stochastic process is given by
\begin{equation}
\rho_t^{(Z_t, \CP)} = \Tr_{\CP} \prtg{\sigma_t^{(Z_t, \CP)}}.
\end{equation}
However, we remark that its statistical average can be simply computed as:
\begin{equation}
\rho_t^{(\CP)} = \Tr_{\CP} \prtg{\mcU_\CP (t) \sigma_0},
\end{equation}
without reference to the spontaneous measurements stochastic process. Indeed, this happens because tracing over the CP-qubits is exactly the same as performing any projective measurements on them and discarding the results.

Let us now introduce $\rho_0$, i.e., the reduced state of system $S$ at time $t=0$. We are interested in obtaining $\rho_t$, that is, the statistical density operator of quantum matter under the assumption that $\rho_0$ is the reduced state of system $S$ at time $t=0$.
We start by noticing that we can always decompose the initial state $\sigma_0$ as
\begin{equation}\label{eq:InitialStateFactorization}
\sigma_0 = \rho_0 \otimes \Omega_\CP + \chi_0,
\qquad
\chi_0 := \sigma_0 - \rho_0 \otimes \Omega_\CP,
\end{equation}
where $\Omega_\CP$ denotes the state in which all CP-qubits are in $\ket{0}$ and $\chi_0$ encodes both the initial correlations between quantum matter and the CP-qubits as well as deviations of the reduced state of the CP-qubits from $\Omega_\CP$.
Since we do not know the disposition of the CPs, nor the actual total initial state $\sigma_0$, one may wonder if $\rho_0$ is the initial state of quantum matter that should be used for making predictions. However, the point is that we \emph{assume} that $\rho_0$ is the initial (reduced) state of quantum matter so that whatever the disposition of the CPs and the initial state $\sigma_0$, one has that $\Tr_{\CP}\prtg{\sigma_0}=\rho_0$. Then, given an initial state $\rho_0$, the statistical final state of quantum matter at time $t>0$ is given by
\begin{multline}
\rho_{t} = \Tr_{\CP}\prtg{\mbE_{\CP}\prtq{
\mcU_{\CP} (t)(\rho_0\otimes \Omega_{\CP})}}
+\\+
\Tr_{\CP}\prtg{\mbE_{\CP}\mbE_{\chi_{0}}\prtq{\mcU_{\CP} (t)\chi_{0}}},
\end{multline}
where $\mbE_{\CP}$ denotes the average over the Poisson sprinkling, while $\mbE_{\chi_0}$ denotes the average over the operators $\chi_0$ that are consistent with the initial state $\rho_0$. The characterization of the distribution of the $\chi_0$s compatible with $\rho_0$ is outside the scope of this paper and we will later argue why we assume that their contribution can be neglected when considering $t \gg \tau_C$.
To proceed with our derivation, we assume that this is the case. Therefore, our focus will be the map defined by
\begin{equation}
\Phi_{t}
:=
\Tr_{\CP}\prtg{\mbE_{\CP}\prtq{
\mcU_{\CP} (t) (\bullet \otimes \Omega_{\CP})}}.
\end{equation}

To find a nicer expression for the map $\Phi_{t}$ we consider the following fact: tracing over the CP-qubits is equivalent to measuring them in any basis and averaging over the possible results. Thus, we can choose to \enquote{measure} the observable $\hX$ for each CP-qubit, the advantage being that $\comm{\hH (t)}{\hX_n}=0$ for all CP-qubits. Therefore, this measurement can be done even before evolving the state. Starting from $\rho_0 \otimes \Omega_{\CP}$, this measurement gives $\pm 1$ for each CP-qubit with $50\%$ probability and all outcomes are independent of each other. Thus, we obtain that
\begin{equation}\label{eq:CPQM_UnitaryUnraveling}
\Phi_{t}
=
\mbE_{\CP}\mbE_{X} \prtq{\mcU_{\CP}^{(X)} (t)},
\end{equation}
where $X$ represents the total string of outcomes of the measurement $\hX$ for each CP-qubit and $\mcU_{\CP}^{(X)} (t)$ is the unitary map acting on quantum matter according to the outcome $X$ and the CPs location, i.e., it is generated by the Hamiltonian $\sqrt{\Upsilon}\hH^{(X)} (t)$, with
\begin{equation}\label{eq:StochasticHamiltonianCPQM}
\hH^{(X)} (t) = \sum_n X_n \tL_{x_n} (t),
\end{equation}
where $x_n$ runs over all CPs locations, each $X_n = \pm 1$ with $50\%$ probability, and all outcomes $X_n$ are independent of each other.

The question remains open: whether for $t\gg \tau_C$, one gets $\rho_t \simeq \Phi_t \rho_0$ and the equality in the limit $t \to \infty$.
Alternatively, one could ask if it is reasonable to expect that the initial dependence on $\chi_0$ becomes negligible for observables evaluated at times $t \gg \tau_C$. In the following, we argue why this expectation is reasonable.

In general, if any initial state $\sigma_0$ is allowed, the above expectation would probably not be true. However, the initial state $\sigma_0$ comes from a previous evolution and is therefore not completely arbitrary. In particular, if one assumes that the flash probability is related to the average value of a smeared (on a spatial radius $r_C$) spatially localized operator $\hO_{r_C}$, such as the (smeared) particle number or the (smeared) mass density, the only CP-qubits that can have a non-negligible amplitude in $\ket{1}$ are those for which $\ev{\hO_{r_C}}_t$ has not been negligible for $t<0$ and $\abs{t}\lesssim \tau_C$. For example, consider a single-particle system which at $t=0$ is on average at rest and located around $\bx_0$; it would be quite unnatural to consider an initial state $\sigma_0$ with excited CP-qubits with $0<t_C \lesssim \tau_C$ and $\abs{\bx_C-\bx_0} \gg r_C$. The only non-negligibly excited CP-qubits with $ 0 < t_C \lesssim \tau_C$ should be those with $\abs{\bx_C-\bx_0} \lesssim r_C$. Moreover, one expects that the flash probabilities for $t_C \gg \tau_C$ will be concentrated around $\bx_0$. Generalizing from the example, we conclude that the sequence of flashes has to show some kind of consistency with the initial state, and this leads us to expect that, in the long run, one has $\rho_t = \Phi_t \rho_0$.

The fact that we can ignore $\chi_0$ in the long run can also be seen from the perspective of non-conserved quantities. Spontaneous collapse models are diffusive~\cite{Donadi2023DiffusiveCollapse}, thus leading to energy increasing over time. If the initial system has a finite amount of energy (e.g., when one considers a finite number of particles), the energy increase due to the spontaneous collapse dynamics must be finite as well over any bounded time interval. From the point of view of the true dynamics, the CPs with $t_C>0$ and $t_C \lesssim \tau_C$ have nothing special and therefore each flash should contribute (statistically) to the energy increase in the same way as those with $t_C \gg \tau_C$. In the long run, flashes with $0 < t_C \lesssim \tau_C$ will be a small part of the total, again leading to the idea that $\lim_{t \to \infty} \rho_t = \lim_{t \to \infty} \Phi_t \rho_0$.

\section{Dynamical evolution in the PSL limit\label{Sec:EvolutionPSL_Limit}}

We are interested now in studying the PSL limit which, as we recall, consists of the following sets of limits:
\begin{equation}\label{eq:PSL_limit}
\Upsilon \to 0,
\quad 
\mu_C \to \infty,
\quad 
\Upsilon \mu_C \to \lambda >0.
\end{equation}
We start by dividing spacetime into cells and number them. For the moment, we keep the size $(\Delta r)^4$ of these cells finite so that when $\mu_C \to \infty$ there will be a divergent number of CPs within each cell.
Eventually, we will take the limit $(\Delta r)^4 \to \dd[4]{r}$.\footnote{Importantly, the limit $(\Delta r)^4 \to \dd[4]{r}$ is taken \emph{after} the limit $\mu_C \to \infty$.}
Moreover, as the size of the cells is arbitrarily small, we can set $\tL_{x_n}=\tL_{x_m}$ when $x_n$ and $x_m$ belong to the same cell. The sum of all outcomes of the $\hX$ measurement within a cell becomes a random walk variable in the limit of infinite steps, i.e., a Gaussian variable with zero mean and variance $\mu_C (\Delta r)^4$~\cite{Book_Oksendal2003StochasticDifferentialEquations,Schondorf2019WienerMeasureDonsker,Book_Durrett2019ProbabilityTheoryExamples}. So, our stochastic Hamiltonian becomes
\begin{gather}
\sqrt{\Upsilon}\hH^{(X)}(t)
\to
\sqrt{\lambda}\hH^{(W)} (t),
\nonumber
\\
\hH^{(W)} (t):= \sum_n \Delta W_n \tL_{x_n} (t),
\label{eq:StochasticHamiltonian_GaussianPSL}
\end{gather}
where
\begin{equation}
\Delta W_n = \sqrt{\Upsilon/\lambda}\sum_{m} X_m,
\qquad
\text{for $m\in n$-th cell},
\end{equation}
and $\Delta W$ are uncorrelated Gaussian variables such that
\begin{equation}
\mbE_W [\Delta W_n]=0,
\qquad
\mbE_W [\Delta W_n \Delta W_m]= \delta_{n,m} (\Delta r)^4.
\end{equation}

Through this cell decomposition, we obtained the Hamiltonian of Eq.~\eqref{eq:StochasticHamiltonian_GaussianPSL}, which provides an unraveling equivalent to that of Eq.~(18) of Ref.~\cite{Diosi2014NonMarkovianGaussianDynamics}.
In Ref.~\cite{Diosi2014NonMarkovianGaussianDynamics}, they study the stochastic equation
\begin{equation}\label{eq:DiosiFerialdiStochasticEquation}
\dv{t}\ket{\psi_t} = -i \hA_t^{j}\phi_j (t)\ket{\psi_t},
\quad
\mbE\prtq{\phi_j (\tau)\phi_k(s)} = D_{jk} (\tau,s),
\end{equation}
where the Einstein's summation convention is employed and the $\phi_j (\tau)$ are real Gaussian noises with zero average. Moreover, Ref.~\cite{Diosi2014NonMarkovianGaussianDynamics} provides the formal map quantum channel $\mcM_t$ that gives the statistical evolution of the density operator according to the stochastic evolution of Eq.~\eqref{eq:DiosiFerialdiStochasticEquation}.
We can identify our problem with the one studied in Ref.~\cite{Diosi2014NonMarkovianGaussianDynamics} by the identification
$\mbE[(\Delta W)_j (\Delta W)_k] = D_{jk} (\tau,s) = \delta_{j,k} (\Delta r)^4$ and immediately get a closed form for $\Phi_t$.
Thus, taking the limit $(\Delta r)^4 \to \dd[4]{r}$, Eq. (17) of Ref.~\cite{Diosi2014NonMarkovianGaussianDynamics} provides the map
\begin{equation}\label{eq:ExactNonMarkovianMap}
\Phi_{t} 
= 
\mcT_{\tmcL} \exp{-\frac{\lambda}{2}\int \dd[4]{r} \int_{0}^{t} \dd{\tau} \int_{0}^{t} \dd{s} \tmcL_{r} (\tau) \tmcL_{r} (s)},
\end{equation}
where $\tmcL_{r} (t) := \comm{\tL_r (t)}{\bullet}$. We remark that the time-ordering here acts on time labels $\tau$ and $s$ but not on $r^0$. Indeed, if we now set $\hL_{x_C} (t) \propto \delta(t-t_C)$ we get
\begin{equation}
\Phi_{t} 
=
\mcT_{\tmcL} \exp{-\frac{\lambda}{2}\int \dd[3]{\bx} \int_{0}^{t} \dd{\tau} \prt{\tmcL_{(\tau,\bx)} (\tau)}^2},
\end{equation}
which leads, in \Schr picture, to the Markovian master equation
\begin{equation}\label{eq:MarkovianMasterEquationGeneral}
\dot{\rho}_t = -i\comm{\hH_S}{\rho_t} -\frac{\lambda}{2} \int \dd[3]{\bx}\comm{\hL_{\bx}}{\comm{\hL_{\bx}}{\rho_t}},
\end{equation}
i.e., the PSL or CSL Markovian master equation~\cite{Piccione2025ExploringMassDependence}.

\subsection{Connection with non-Markovian CSL}

We can now connect our equations with non-Markovian collapse models without jumps~\cite{Adler2007NonWhiteNoises,Adler2008NonWhiteNoisesII,Diosi2014NonMarkovianGaussianDynamics,Gundhi2025RelativisticCollapse}. We take into consideration operators $\hL_x (t)$ of the following form:
\begin{equation}\label{eq:CollapseOperatorCSL}
\hL_x (t) = \int \dd[3]{\by} g (x^0-t;\bx-\by) \hA_{\by},
\end{equation}
where $\hA_\by$ is a time-independent Hermitian operator and $g(t;\bx)$ is a generic function with $g(t;\bx) \in \mathbb R$, $g(t;\bx)=0$ for $t<0$. A common choice is to take $\hA_\by$ as the mass density operator $\hmu$ at point $\by$, i.e. $\hA_\bx = \hmu (\bx)$. By now dividing the Hamiltonian of Eq.~\eqref{eq:StochasticHamiltonian_GaussianPSL} by $(\Delta r)^4$ and taking the limit $\Delta r \to \dd{r}$ we get\footnote{$w(r):= \lim_{\Delta r\to 0} \Delta W_n/(\Delta r)^4$ is a white noise in spacetime. Thus, one has $\mbE[w(x)w(y)]=\delta(x-y)$.}
\begin{multline}\label{eq:StochasticHamiltonianCSLUnraveling}
\hH^{(W)} (t) 
= 
\sum_n (\Delta r)^4 \frac{\Delta W_n}{(\Delta r)^4} \tL_{x_n} (t)
\to\\
\to
\int \dd[4]{r} \dd[3]{\by} \tl A_{\by} (t) g (r^0-t;\br-\by) w(r)
=\\
=
\int \dd[3]{\by} \phi_\by (t) \tl A_{\by} (t),
\end{multline}
where $w(r)$ is a white noise in spacetime, $\tl A_{\by} (t)$ corresponds to $\hA_\by$ in the interaction picture, and (by identifying $t$ with $y^0$) $\phi_\by (t)$ is a Gaussian noise defined as
\begin{equation}
\phi_\by (t) 
:= \int \dd[4]{r} g (r^0-t;\br-\by) w(r) 
= \int \dd[4]{r} g (r-y) w(r).
\end{equation}
It follows that
\begin{equation}
\mbE[\phi (x)]=0,
\quad
\mbE[\phi (x) \phi (y)] = G(x,y),
\end{equation}
where
\begin{equations}\label{eq:NoiseCorrelationFunction}
G(x,y)
:&= \int \dd[4]{r} g (r-x) g (r-y),
\\
&= \int \dd[4]{r} g (r) g (r+(x-y)),
\end{equations}
i.e., $G(x-y)$ is the autocorrelation function of $g(x-y)$. Importantly, we assumed [cf. the text below Eq.~\eqref{eq:InteractionHamiltonianStructure}] that $g(t;\bx)$ is characterized by a time $\tau_C$ such that $g(t;\bx)\sim 0$ when $t \gg \tau_C$. It follows that $G(x^0 - y^0;\bx-\by)$ is negligible when $\abs{x^0-y^0} \gg \tau_C$. In other words, $\tau_C$ is basically the correlation time of the noise $\phi_\bx (t)$.

Substituting Eq.~\eqref{eq:CollapseOperatorCSL} in Eq.~\eqref{eq:ExactNonMarkovianMap} we get\footnote{One can also take the stochastic Hamiltonian of Eq.~\eqref{eq:StochasticHamiltonianCSLUnraveling} and take Eq.~\eqref{eq:ExactNonMarkovianMap_CSLOperators} from Ref.~\cite{Diosi2014NonMarkovianGaussianDynamics}.}
\begin{equation}\label{eq:ExactNonMarkovianMap_CSLOperators}
\Phi_{t}
= \mcT_{\tmcA} \exp{-\frac{\lambda}{2}\int_{0}^{t} \dd[4]{y}\dd[4]{z} G(y-z) \tmcA (y) \tmcA (z)},
\end{equation}
where $\tmcA (y) := \comm{\tl{A}_{\by} (y^0)}{\bullet}$, and $\int_{0}^{t} \dd[4]{y}\dd[4]{z}$ means that both $y^0$ and $z^0$ go from $0$ to $t$.

\subsection{Connection with mass-proportional PSL}

In the Markovian PSL model of Refs.~\cite{Piccione2025ExploringMassDependence,Piccione2025NewtonianPSL,Piccione2026EnergyIncreaseGPSL}, the collapse operator $\hL_x (t)$ reads
\begin{equations}\label{eq:MassProportionalCollapseOperatorsPSL}
\hL_x (t) &=
\delta(x^0-t) \sqrt{\hmu_{r_C} (\bx)},
\\
\hmu_{r_C} (\bx) &:= \int \dd[3]{\by} g_{r_C}(\bx-\by)\hmu(\by),
\end{equations}
where $\hmu(\by)$ is the mass density operator and $g_{r_C} (\bx)$ is a smearing distribution characterized by a length $r_C$. This makes the flash probability proportional to the smeared mass density $\hmu_{r_C} (\bx)$ and allows the introduction of semiclassical Newtonian gravity as a feedback mechanism on top of the spontaneous collapse one~\cite{Piccione2023Collapse,Piccione2025NewtonianPSL,Piccione2026EnergyIncreaseGPSL}.
Indeed, substituting $\hL_{\bx}$ with $\sqrt{\hmu_{r_C} (\bx)}$ in Eq.~\eqref{eq:MarkovianMasterEquationGeneral} gives the PSL master equation studied in Refs.~\cite{Piccione2025ExploringMassDependence,Piccione2025NewtonianPSL,Piccione2026EnergyIncreaseGPSL}.
A natural way to get a non-Markovian extension of such models is to set
\begin{equation}
\hL_x (t) =
f_{\tau_C} (x^0-t)\sqrt{\hmu_{r_C} (\bx)},
\qquad
\int \dd{t} f_{\tau_C} (t) = 1,
\end{equation}
where $f_{\tau_C} (t)=0$ for $t<0$ and $\tau_C$ basically sets for how long a CP-qubit interacts with quantum matter. 

\section{supercumulant expansion and TCL master equation\label{Sec:TCLMasterEquation}}

In the previous section, we derived Eq.~\eqref{eq:ExactNonMarkovianMap}, which is, however, difficult to use for calculations. We can derive a time-local master equation by exploiting the expansion in supercumulants~\cite{Szankowski2023OpenQuantumSystems,Szankowski2025OpenQuantumSystems}.\footnote{Ref.~\cite{Szankowski2025OpenQuantumSystems} is the updated version of Ref.~\cite{Szankowski2023OpenQuantumSystems}.} This is done by writing the map of Eq.~\eqref{eq:ExactNonMarkovianMap} as follows
\begin{equation}\label{eq:SuperCumulantExpansion}
\Phi_{t} = \mcT_\mcC \exp{\sum_{k=1}^\infty \prt{-i\sqrt{\lambda}}^k \int_{0}^{t} \mcC_k (s) \dd{s}},
\end{equation}
where $\mcT_\mcC$ is now a time-ordering at the ``coarse-grained'' level of the supercumulants, i.e. one has that
\begin{equation}
\mcT_\mcC \prtg{\mcC_k (t)\mcC_{k'} (s)}
=
\begin{cases}
\mcC_k (t)\mcC_{k'} (s), \quad & t\geq s,
\\
\mcC_{k'} (s)\mcC_k (t), \quad & t < s.
\end{cases}
\end{equation}
As shown in Ref.~\cite{Szankowski2025OpenQuantumSystems}, the convenience of this supercumulant expansion is that by truncating at a given order $n$, one has
\begin{equation}
\dv{t} \Phi_{t}^{(n)} = \prt{\sum_{k=1}^n \prt{-i\sqrt{\lambda}}^k \mcC_k (t)}\Phi_{t}^{(n)},
\end{equation}
thus allowing us to derive a time-local master equation, the so-called TCL (Time-ConvolutionLess) master equation of order $n$~\cite{Book_Breuer2002}.

When the bath is Gaussian, as in our case, every odd supercumulant is vanishing~\cite{Szankowski2025OpenQuantumSystems}.
The contribution of each even-order supercumulant to the dynamics up to time $t$ can be estimated as\footnote{See section 4.6, page 51, Eq.~(238), of~\cite{Szankowski2025OpenQuantumSystems}. The operator norm used in the equation may be problematic for unbounded operators. However, to make physical estimations, one can simply substitute the operator norm with the expected order of magnitude of expectation values of the operators.}:
\begin{equation}
\norm{\lambda^{k}\int_0^t \mcC_{2k} (s) \dd{s}}
\sim
t\sqrt{\lambda} \prt{\tau_C\sqrt{\lambda}}^{2k-1}\prt{\frac{\tau_C}{\tau_S}}^{k-1},
\end{equation}
where $\tau_C$ is the correlation time of the noise and $\tau_S$ (assumed to be much higher than $\tau_C$) is the shortest time-scale associated with the free dynamics of our system of interest. 
Thus, the relative contribution of two supercumulants (order $k$ and $n$ respectively) gives
\begin{equation}
\frac{\norm{\lambda^{k}\int_0^t \mcC_{2k} (s) \dd{s}}}{\norm{\lambda^{n}\int_0^t \mcC_{2n} (s) \dd{s}}}
\sim
\prt{\lambda\frac{\tau_C^3}{\tau_S}}^{k-n}.
\end{equation}
When $\lambda \ll \tau_S/\tau_C^3$, one can safely stop at the first non-vanishing supercumulant for \emph{any} time $t$, even for $t \gg \tau_C$.

Following Ref.~\cite{Szankowski2025OpenQuantumSystems}, by expanding Eq.~\eqref{eq:ExactNonMarkovianMap} and Eq.~\eqref{eq:SuperCumulantExpansion} to the same order in $\lambda$ we get 
\begin{equation}\label{eq:TCL_SecondOrder}
\mcC_2 (t) = \int \dd[4]{r} \int_{0}^{t} \dd{s} \tmcL_{r} (t) \tmcL_{r} (s).
\end{equation}
Assuming that $\hL_{x_C} (t) \propto \delta(t-t_C)$, Eq.~\eqref{eq:TCL_SecondOrder} reduces to Eq.~\eqref{eq:MarkovianMasterEquationGeneral}, i.e., the Markovian master equation.\footnote{One should keep in mind that $\int_{0}^{t} \delta (t-s)f(s)\dd{s} = f(t)/2$.}


By specializing Eq.~\eqref{eq:TCL_SecondOrder} to the case of Eq.~\eqref{eq:CollapseOperatorCSL}, we get
\begin{equation}
\mcC_2 (t)
=
\int \dd[3]{\bx}\dd[3]{\by}
\int_{0}^t \dd{s}
G(t-s;\bx-\by)\tmcA_{\bx} (t)\tmcA_{\by} (s).
\end{equation}
This leads to a TCL master equation that is exactly the one in Eq.~(39) of Ref.~\cite{Adler2007NonWhiteNoises},\footnote{Eq.~(39) of Ref.~\cite{Adler2007NonWhiteNoises} is \Schr picture while ours is given in interaction picture. Moreover, to see the equivalence, one must identify our $\lambda$ with the term $\gamma\abs{\xi}^2$ of Ref.~\cite{Adler2007NonWhiteNoises}, and make continuous the equation in Ref.~\cite{Adler2007NonWhiteNoises} or discretize the spatial integrals in ours.} i.e.:
\begin{equation}
\dot{\rho}_t
=
-\lambda \mcC_2 (t) \rho_t.
\end{equation}

\section{Flash probabilities\label{Sec:FlashProbabilities}}

\subsection{General case}

Our final task consists of characterizing the statistics of the flash process. Given the initial matter state
$\rho_0$ at $t=0$, and working in the regime $t_C \gg \tau_C$, we denote
by $\mu_F^{(n)}(x_1,\dots,x_n|\rho_0)$ the $n$-point joint intensity density
of the flash process. Thus,
$\mu_F^{(n)}(x_1,\dots,x_n|\rho_0)\dd[4]{x_1}\cdots\dd[4]{x_n}$
is the probability, to leading order in the infinitesimal four-volumes, that
one flash occurs in each of the neighborhoods $\dd[4]{x_i}$ of the spacetime
points $x_i$, irrespective of possible flashes elsewhere. Unlike a Poisson point process, the flash process is not entirely determined by its
first-order intensity density $\mu_F^{(1)}(x|\rho_0)$.\footnote{For a Poisson
point process, all factorial moment densities factorize into products of the
first-order intensity density. In our case this is generally not true. For
instance, if a single particle is prepared in a superposition of two distant
wave packets, the occurrence of a flash near one packet changes the conditional
state and thereby suppresses the probability of subsequent flashes near the
other packet. Hence, in general,
$\mu_F^{(2)}(x_1,x_2|\rho_0)\neq
\mu_F^{(1)}(x_1|\rho_0)\mu_F^{(1)}(x_2|\rho_0)$.}
A complete characterization therefore requires the full set of the joint
intensity densities $\mu_F^{(n)}(x_1,\ldots,x_n|\rho_0)$, for all $n\geq 1$. In Appendix~\ref{APPSec:PointProcessMathematics}, we summarize why the knowledge of all the $\mu_F^{(n)}(x_1,\ldots,x_n|\rho_0)$ completely characterize the stochastic flash process.

The probability $\mu_F^{(n)}(x_1,\dots,x_n|\rho_0)\dd[4]{x_1}\cdots\dd[4]{x_n}$ can be computed as
\begin{multline}
\mu_F^{(n)}(x_1,\dots,x_n|\rho_0)\dd[4]{x_1}\cdots\dd[4]{x_n}
=\\
=
P (\textrm{Flashes}|\CP,\rho_0) \mu_\CP(x_1,\dots,x_n)\dd[4]{x_1}\cdots\dd[4]{x_n},
\end{multline}
where $\mu_\CP(x_1,\dots,x_n)$ is the $n$-point intensity for the Poisson sprinkling:
\begin{equation}
\mu_\CP(x_1,\dots,x_n)\dd[4]{x_1}\cdots\dd[4]{x_n}
=
\mu_C^n \dd[4]{x_1}\cdots\dd[4]{x_n}.
\end{equation}
Thus, we now need to compute the conditional probability $P (\textrm{Flashes}|\CP,\rho_0)$ of projecting all $n$ CP-qubits onto $\ket{1}$ given the position of $n$ CPs. Starting from the initial state $\sigma_{0}$, this probability is given by 
\begin{equation}
P (\textrm{Flashes}|\CP,\rho_0)
=
\Tr{\prt{\bigotimes_{j=1}^n \dyad{1}_j}\sigma_{T}},
\end{equation}
where $\sigma_{T}$ is the state of quantum matter plus all the CPs with $t_C>0$ at time $T$, with $T$ greater than all the times $t_C$ of the $n$ CPs we consider. Thus, the probability of having flashes at locations $(x_1,\dots,x_n)$ is given by
\begin{multline}
\mu_F^{(n)}(x_1,\dots,x_n|\rho_0)\dd[4]{x_1}\cdots\dd[4]{x_n}
=\\
=
\mu_C^n \dd[4]{x_1}\cdots\dd[4]{x_n}\Tr{\prt{\bigotimes_{j=1}^n \dyad{1}_j}\sigma_{T}},
\end{multline}

Since we are interested in flashes at points with $t_C \gg \tau_C$, we can assume that\footnote{To be precise, $\Omega_\CP$ that appears in Eq.~\eqref{eq:InitialStateFlashProbabilities} refers to all CP-qubits with $t_C >0$ but not those enumerated by $j=1,\dots,n$.}
\begin{equation}\label{eq:InitialStateFlashProbabilities}
\sigma_T = \mcU (T) \prt{\sigma^{(n)}_{0}\otimes \Omega_\CP},
\qquad
\sigma^{(n)}_{0}:= \rho_0 \otimes_{j=1}^n \dyad{0}_j,
\end{equation}
which basically amounts to neglecting again the initial correlations given by $\chi_0$ in Eq.~\eqref{eq:InitialStateFactorization}. A consistency argument similar to the one we made at the end of Sec.~\ref{Sec:UnitaryUnravelingCPQM} can be made to argue for the validity of this assumption.

\subsection{PSL limit}

Under the assumption that Eq.~\eqref{eq:InitialStateFlashProbabilities} holds, by repeating the same kind of limits performed in Sec.~\ref{Sec:EvolutionPSL_Limit} we get that the interaction picture evolution with respect to $\hH_S$ is given by
\begin{equation}
\hW_{n} (t)
:=
\sqrt{\Upsilon}\hH^{(n)} (t) + \sqrt{\lambda}\hH^{(W)} (t),
\end{equation}
where
\begin{equation}\label{eq:HamiltonianDecompositionForFlashProbabilities}
\hH^{(n)}:= \sum_{j=1}^{n} \tL_{x_j} (t) \hX_j,
\qquad
\hH^{(W)}:= \sum_{k>n} \Delta W_k \tL_{x_k} (t).
\end{equation}
Moreover, we decompose the density matrix $\rho_0$ in a diagonal basis\footnote{
If system $S$ is finite-dimensional (say with dimensions $d$), one can always diagonalize $\rho_0$ so that $\rho_0 = \sum_{j=1}^{d} \kappa_j\dyad{\psi_j}$, with $\kappa_j \geq 0$ and $\sum_{j=1}^{d} \kappa_j = 1$. If system $S$ is infinite-dimensional, one can always arbitrarily well approximate it with such a sum of pure state density matrices.} so that we can focus on an initial pure state $\ket{\psi_0}$. Thanks to the unraveling, we have that
\begin{multline}\label{eq:MultiPointProbabilityFlashesPSL}
\mu_F^{(n)} (x_1,\dots,x_n|\psi_0)
=\\
\mu_C^n \mbE_W\! \prtq{
\Tr{\! \mel{1_n}{U^{(n,W)}}{\psi_0,0_n}
\!\!
\mel{\psi_0,0_n}{U^{(n,W)\dg}}{1_n}}},
\end{multline}
where $\ket{0_n}:=\ket{0}_1\cdots\ket{0}_n$, $\ket{1_n}:=\ket{1}_1\cdots\ket{1}_n$, and
\begin{equation}\label{eq:UnitaryEvolutionFlashesPSL}
U^{(n,W)}
:=
\mcT_{\hW_{n}} \exp{-i \int_{0}^{T}\dd{s} \hW_{n} (s)}.
\end{equation}
When expanding the above operator, one should compute quantities such as
\begin{equation}\label{eq:ExpansionUnitaryFlashIntensities}
\frac{(-i)^m}{m!}\bra{1_n}\mcT_{\hW_n}
\prtg{\hW_n(s_1)\cdots\hW_n(s_m)}\ket{0_n}.
\end{equation}
Therefore, in the expansion, all terms of order $m<n$ vanish because every ket $\ket{0}$ has to be flipped to $\ket{1}$. So, both factors inside the trace contribute with terms that are at least of order $\Upsilon^{n/2}$. At the same time, since we are interested in the PSL limit, every term that will lead, overall, to terms of order higher than $\Upsilon^{n/2}$ will vanish. Thus,
we find that the contributing terms are (a more detailed and explanatory calculation is provided in Appendix~\ref{APPSec:FlashIntensityDetailedCalculation})
\begin{equation}
\mel{1_n}{U^{(n,W)}}{\psi_0,0_n}
\to
\Upsilon^{n/2}\hK^{(n,W)}\ket{\psi_0},
\end{equation}
where
\begin{multline}
\hK^{(n,W)}
:=\\
\sum_{m=n}^{\infty}
\frac{(-i)^m \lambda^{(m-n)/2}}{m!}
\!\int_0^T\! \dd{s_1} \cdots \dd{s_m} \hM_{n,m} (s_1,\dots,s_m),
\end{multline}
where\footnote{We point out that $\hM_{n,m}$ does not contain factors of $\lambda$ after the $\Upsilon\to 0$ limit. We refer the reader to Appendix~\ref{APPSec:FlashIntensityDetailedCalculation} for more details on the combinatorial structure of $\hM_{n,m}$.}
\begin{multline}
\hM_{n,m} (s_1,\dots,s_m)
:=\\
\lim_{\Upsilon\to 0}
\frac{\bra{1_n}\mcT_{\hW_n}
\prtg{\hW_n(s_1)\cdots\hW_n(s_m)}\ket{0_n}}{\Upsilon^{n/2}\lambda^{(m-n)/2}}.
\end{multline}

Thus, the final result in the PSL limit is
\begin{equation}\label{eq:FlashProbabilities}
\mu_F^{(n)} (x_1,\dots,x_n|\rho_0)
=
\lambda^n \mbE_W \Tr{\hK^{(n,W)}\rho_0\hK^{(n,W)\dg}}.
\end{equation}
This shows that the PSL limit can be consistently taken at the level of late-time flash probabilities ($t_1,\dots,t_n \gg \tau_C$), and we found the formula to compute the intensities of all orders.

To understand how these probabilities behave, we can assume, as commonly done in spontaneous collapse models~\cite{Adler2007Bounds,Figurato2024DPEffectiveness}, that $\hH_S$ is negligible. Moreover, we assume that $\comm{\hL_{x_n} (t)}{\hL_{x_m} (s)} =0$. This happens, for example, by employing the collapse operators given in Eq.~\eqref{eq:MassProportionalCollapseOperatorsPSL}. In this case, we have [cf. Eq.~\eqref{eq:UnitaryEvolutionFlashesPSL}]
\begin{multline}
U^{(n,W)}
:=
\prod_{j=1}^{n} \exp{-i \sqrt{\Upsilon}\int_{0}^{T} \dd{s} \hL_{x_j} (s) \hX_j}
\\
\prod_{k>n} \exp{-i \sqrt{\lambda}\int_{0}^{T} \dd{s} \Delta W_k \hL_{x_k} (s)},
\end{multline}
so that, inserting this into Eq.~\eqref{eq:MultiPointProbabilityFlashesPSL}, one gets that the noise-dependent part cancels. Thus, one gets
\begin{equation}
\mu_F^{(n)} (x_1,\dots,x_n|\rho_0)
=
\lambda^n \Tr{\hK^{(n)}\rho_0\hK^{(n)\dg}},
\end{equation}
where
\begin{equation}
\hK^{(n)}
:=
(-i)^n
\!\int_0^T\!\! \dd{s_1} \cdots \dd{s_n} \hL_{x_1} (s_1) \cdots \hL_{x_n} (s_n).
\end{equation}
Considering a $T\gg \tau_C$ and a set of points $(x_1,\dots,x_n)$ with $t_C \gg \tau_C$, we find that the above quantity can also be computed as
\begin{equation}
\hK^{(n)}
=
(-i)^n
\!\intmp \!\!\! \dd{s_1} \cdots \dd{s_n} \hL_{x_1} (s_1) \cdots \hL_{x_n} (s_n).
\end{equation}
On the other hand, considering a single flash probability $\mu_F^{(1)} (x|\rho_0)$ with $t_C \ll \tau_C$, we get $\mu_F^{(1)} (x|\rho_0) \sim 0$, which is clearly nonsensical.
This shows that the contribution of $\chi_0$ does not vanish uniformly even in the PSL limit: initial correlations between system $S$ and the CP-qubits are needed to obtain the correct short-time flash probabilities.

\section{Discussions and conclusions\label{Sec:Conclusions}}


We have developed a non-Markovian extension of the PSL model~\cite{Piccione2023Collapse,Piccione2025ExploringMassDependence} by allowing each Collapse Point (CP) to interact with quantum matter on a finite time scale $\tau_C$. The Markovian model is indeed recovered for $\tau_C \to 0$. We first presented the model with a finite CP density $\mu_C$. We associated with each CP a qubit (a CP-qubit) that interacts with standard quantum matter, with coupling strength $\sqrt{\Upsilon}$. We then studied the PSL limit: $\mu_C \to \infty$, $\Upsilon \to 0$, $\mu_C\Upsilon \to \lambda >0$.

Our first main result consists of having identified the map $\Phi_t$ in the PSL limit [Eq.~\eqref{eq:ExactNonMarkovianMap}]. Under the physically motivated assumption that the initial correlations between quantum matter and CP-qubits become negligible at late times, $\Phi_t \rho_0$ gives the state of quantum matter at time $t \gg \tau_C$ given the state $\rho_0$ at time $t=0$. Our second main result is the derivation of the corresponding time-convolutionless master equation through the supercumulant expansion~\cite{Szankowski2023OpenQuantumSystems,Szankowski2025OpenQuantumSystems}. Stopping the expansion at the first non-vanishing order and choosing a specific form of interaction between CP-qubits and standard quantum matter allows us to recover known master equations in the non-Markovian spontaneous collapse models literature~\cite{Adler2007NonWhiteNoises,Gundhi2025RelativisticCollapse}. Finally, given $\rho_0$ at time $t=0$, we derived the flash intensities (of all orders) at given spacetime points with $t_C \gg \tau_C$. This shows that the PSL limit can also be taken consistently at the level of flash statistics.

A central conceptual point is the role of the initial correlations encoded in $\chi_0$ [see Eq.~\eqref{eq:InitialStateFactorization}]. Even in the PSL limit, these correlations cannot in general be neglected at short times: doing so leads to unphysical conclusions for the flash intensities when considering spacetime points with time-coordinate $t_C$ such that $0 < t_C \lesssim \tau_C$. This is consistent with the general fact that reduced dynamics can depend crucially on initial system-environment correlations, especially in the non-Markovian regime~\cite{Pechukas1994ReducedDynamicsNeedNotBeCompletelyPositive,Review_deVega2017DynamicsNonMarkovianOpenQuantumSystems,Book_Breuer2002}. By contrast, the factorized map $\Phi_t$ is expected to emerge as an effective long-time description.

Overall, the present framework shows that PSL admits a mathematically well-defined non-Markovian extension, recovers the known Markovian limit, and retains a clear spacetime-flash interpretation~\cite{Piccione2023Collapse,Piccione2025ExploringMassDependence}. For this reason, we believe that it provides a promising starting point for a relativistic extension. In particular, a natural strategy for future work is to search for collapse operators and correlation functions that lead to the statistical dynamics investigated in Ref.~\cite{Gundhi2025RelativisticCollapse}. This will be the subject of future work.

\section*{Acknowledgements}

N.P. acknowledges support from the PNRR MUR projects PE0000023-NQSTI, INFN, and the University of Trieste.
N.P. acknowledges support also from the European Union Horizon’s 2023 research and innovation program [HORIZON-MSCA-2023-PF-01] under the Marie Sk{\l}odowska-Curie Grant Agreement No. 101150889 (CPQM).

N.P. thanks Angelo Bassi for feedback on this paper's draft and discussions about the results.


%

\clearpage
\onecolumngrid
\appendix

\section{Flash intensities completely characterize the stochastic flash process\label{APPSec:PointProcessMathematics}}

In this Appendix we summarize how, and in what sense, the intensities $\mu_F^{(n)}$ of Sec.~\ref{Sec:FlashProbabilities} completely characterize the stochastic flash process. Even though the formulas in Sec.~\ref{Sec:FlashProbabilities} are (approximately) valid only when all flash locations are far in the future (with respect to $\tau_C$), we will here consider $\mu_F^{(n)}(x_1,\dots,x_n|\rho_0)$ to be valid for all $x$ in spacetime. Since the formulas are intended to predict the flash statistics in the future of $t=0$, we assume that, whenever one of the $x_j$ is such that $t_j \leq 0$, we have that $\mu_F^{(n)}(x_1,\dots,x_n|\rho_0)=0$. All the equations presented in this Appendix are taken from chapter 5 of Ref.~\cite{Book_Daley2003PointProcessesVol1}. As this is a physics paper, we assume that the mathematical conditions under which they are derived are satisfied. The general theory of stochastic point processes is presented in Refs.~\cite{Book_Daley2003PointProcessesVol1,Book_Daley2008PointProcessesVol2}.

Given a bounded spacetime region \(A\), one can compute another set of quantities called the local Janossy densities \cite[Def.~5.4.IV, p.~137; Eq.~(5.4.14), p.~138]{Book_Daley2003PointProcessesVol1}:
\begin{equation}
j_A^{(n)}(x_1,\dots,x_n|\rho_0)
=
\sum_{k=0}^{\infty} \frac{(-1)^k}{k!} \int_{A^k} \mu_F^{(n+k)} (x_1,\dots,x_n,z_1,\ldots,z_k|\rho_0) \dd[4]{z_1}\cdots \dd[4]{z_k}.
\end{equation}
where the subscript $A^k$ in the integral indicates that each variable $z_1,\dots,z_k$ should be integrated over the region $A$. The quantity $j_A^{(n)}(x_1,\dots,x_n|\rho_0)\dd[4]{x_1}\cdots\dd[4]{x_n}$ gives the probability that there are exactly $n$ flashes in $A$ at locations $\dd[4]{x_1},\dots,\dd[4]{x_n}$.

We can then introduce the family of stochastic variables $N_A$, which count how many flashes occur in the bounded spacetime region $A$. The probability $\mbP_{\rho_0} (N_A=n)$ of obtaining a certain number $n$ of flashes inside the region $A$ is then given by~\cite[Eq.~(5.3.12), p.~130]{Book_Daley2003PointProcessesVol1}\footnote{The quantity $\sum_{r=0}^{\infty} (r!)^{-1} J_{n+r} \prt{A^{(n)}\times (A^{c})^{(r)}}$ in Eq.~(5.3.12) of Ref.~\cite{Book_Daley2003PointProcessesVol1} corresponds to $\int_{A^n} j_A^{(n)}(x_1,\dots,x_n|\rho_0)\dd[4]{x_1}\cdots\dd[4]{x_n}$ according to \cite[Def.~5.4. IV, p.~137]{Book_Daley2003PointProcessesVol1}.}
\begin{equation}
\mbP_{\rho_0} (N_A=n) 
= 
\frac{1}{n!}\int_{A^n} j_A^{(n)}(x_1,\dots,x_n|\rho_0)\dd[4]{x_1}\cdots\dd[4]{x_n}.
\end{equation}
More generally, for a finite family of disjoint bounded regions
\(A_1,\ldots,A_q\), with \(C:=A_1\cup\cdots\cup A_q\), one has\cite[Eq.~(5.3.13), p.~130]{Book_Daley2003PointProcessesVol1}\footnote{The quantity $\sum_{r=0}^{\infty} (r!)^{-1} J_{n+r} \prt{A_1^{(n_1)}\times \cdots A_q^{(n_q)} \times (C^{c})^{(r)}}$ in Eq.~(5.3.13) of Ref.~\cite{Book_Daley2003PointProcessesVol1} corresponds to $\int_{A_1^{n_1}}\!\!\cdots\!\int_{A_q^{n_q}}
j_C^{(n_1+\cdots+n_q)}
(x_1^{(1)},\ldots,x_{n_q}^{(q)}|\rho_0)
\prod_{\alpha=1}^{q}
\prod_{j=1}^{n_\alpha}\dd[4]{x_j^{(\alpha)}}$ according to \cite[Def.~5.4. IV, p.~137]{Book_Daley2003PointProcessesVol1}.}
\begin{equation}
\mbP_{\rho_0}
\left(
N_{A_1}=n_1,\ldots,N_{A_q}=n_q
\right)
=
\frac{1}{n_1!\cdots n_q!}
\int_{A_1^{n_1}}\!\!\cdots\!\int_{A_q^{n_q}}
j_C^{(n_1+\cdots+n_q)}
(x_1^{(1)},\ldots,x_{n_q}^{(q)}|\rho_0)
\prod_{\alpha=1}^{q}
\prod_{j=1}^{n_\alpha}\dd[4]{x_j^{(\alpha)}}.
\end{equation}
The collection of all such finite-dimensional count distributions completely
characterizes the point process.\footnote{See also pages 26 and 27 of Ref.~\cite{Book_Daley2008PointProcessesVol2}. In particular, Corollary 9.2.IV.}
Hence, under the assumptions stated above,
the intensities $\mu_F^{(n)}$ determine all flash counting probabilities in bounded spacetime regions.

\clearpage
\section{Details on the derivation of the flash intensities\label{APPSec:FlashIntensityDetailedCalculation}}

We start from Eq.~\eqref{eq:ExpansionUnitaryFlashIntensities} of the main text. Taking as an example the case in which $n=2$ and $m=3$, one gets (assuming that $s_1 > s_2 > s_3$):
\begin{multline}
\bra{1_2}\mcT_{\hW_2}\prtg{\hW_2(s_1)\hW_2(s_2)\hW_2(s_3)}\ket{0_2}
=
\bra{1_2}\hW_2(s_1)\hW_2(s_2)\hW_2(s_3)\ket{0_2}
=\\
=
\bra{1_2}
\prt{\sqrt{\Upsilon}\tL_{x_1}(s_1)\hX_1+\sqrt{\Upsilon}\tL_{x_2}(s_1)\hX_2 + \sqrt{\lambda}\hH^{(W)} (s_1)}
\prt{\sqrt{\Upsilon}\tL_{x_1}(s_2)\hX_1+\sqrt{\Upsilon}\tL_{x_2}(s_2)\hX_2 + \sqrt{\lambda}\hH^{(W)} (s_2)}
\times \\ \times
\prt{\sqrt{\Upsilon}\tL_{x_1}(s_3)\hX_1+\sqrt{\Upsilon}\tL_{x_2}(s_3)\hX_2 + \sqrt{\lambda}\hH^{(W)} (s_3)}
\ket{0_2}
=\\
=
\Upsilon\sqrt{\lambda}
\prtB{
\tL_{x_1}(s_1)\tL_{x_2}(s_2)\hH^{(W)} (s_3)
+
\tL_{x_1}(s_1)\hH^{(W)} (s_2)\tL_{x_2}(s_3)
+
\tL_{x_2}(s_1)\tL_{x_1}(s_2)\hH^{(W)} (s_3)
+
\tL_{x_2}(s_1)\hH^{(W)} (s_2)\tL_{x_1}(s_3)
+\\
+
\hH^{(W)} (s_1)\tL_{x_1}(s_2)\tL_{x_2}(s_3)
+
\hH^{(W)} (s_1)\tL_{x_2}(s_2)\tL_{x_1}(s_3)
}.
\end{multline}
If we are not in the case $s_1 > s_2 > s_3$, we can restore the time-ordering at the end because the surviving terms take just one operator from each $\hW_2 (s)$.

Let us consider another example: $n=1$ and $m=3$. To start, we consider again that $s_1>s_2>s_3$ and we obtain
\begin{multline}
\bra{1_1}\mcT_{\hW_1}\prtg{\hW_1(s_1)\hW_1(s_2)\hW_1(s_3)}\ket{0_1}
=
\bra{1_1}\hW_1(s_1)\hW_1(s_2)\hW_1(s_3)\ket{0_1}
=\\
=
\bra{1_1}
\prt{\sqrt{\Upsilon}\tL_{x_1}(s_1)\hX_1 + \sqrt{\lambda}\hH^{(W)} (s_1)}
\prt{\sqrt{\Upsilon}\tL_{x_1}(s_2)\hX_1 + \sqrt{\lambda}\hH^{(W)} (s_2)}
\prt{\sqrt{\Upsilon}\tL_{x_1}(s_3)\hX_1+\sqrt{\lambda}\hH^{(W)} (s_3)}
\ket{0_1}
=\\
=
\sqrt{\Upsilon}\lambda
\prtB{
\tL_{x_1}(s_1)\hH^{(W)}(s_2)\hH^{(W)} (s_3)
+
\hH^{(W)}(s_1)\tL_{x_1}(s_2)\hH^{(W)} (s_3)
+
\hH^{(W)} (s_1)\hH^{(W)}(s_2)\tL_{x_1}(s_3)
}
+\\+
\Upsilon^{3/2}\tL_{x_1}(s_1)\tL_{x_1}(s_2)\tL_{x_1}(s_3).
\end{multline}
Now, as explained in Sec.~\ref{Sec:FlashProbabilities}, the last term with prefactor $\Upsilon^{3/2}$ does not survive the PSL limit, so we can discard it. Then, as in the previous example, if we are not in the case $s_1 > s_2 > s_3$, we can restore the time-ordering at the end because the surviving terms take just one operator from each $\hW_1 (s)$.

Generalizing the previous examples what we understand is the following. First that
\begin{equation}
\frac{(-i)^m}{m!}\bra{1_n}\mcT_{\hW_n}
\prtg{\hW_n(s_1)\cdots\hW_n(s_m)}\ket{0_n}
=
\frac{(-i)^m}{m!}\mcT\prtg{\bra{1_n}\hW_n(s_1)\cdots\hW_n(s_m)\ket{0_n}},
\end{equation}
where $\mcT$ without subscript orders according to the time-argument of all operators appearing inside it. This happens because $\bra{1_n}\hW_n(s_1)\cdots\hW_n(s_m)\ket{0_n}$ will anyway be a sum of operator products in which each factor in the product comes from a different $\hW_n^ (s_j)$. Second, let us imagine to have $m$ slots to fill indexed by $j$ with $j=1,\dots,m$. When $m<n$, Eq.~\eqref{eq:ExpansionUnitaryFlashIntensities} of the main text evaluates to zero. When $m=n$, we must fill every slot $j$ with a different $\tL_{x_k}$, so that there are $n!$ possible sequences. In other words, after the PSL limit, it remains
\begin{equation}
\frac{(-i)^n}{n!}\bra{1_n}\mcT_{\hW_n}
\prtg{\hW_n(s_1)\cdots\hW_n(s_n)}\ket{0_n}
\xrightarrow{\text{PSL limit}}
\frac{(-i)^n}{n!}\Upsilon^{n/2}
\mcT\prtg{\sum_\sigma \prod_{j=1}^{n} \tL_{x_{\sigma^{-1}(j)}} (s_j)},
\end{equation}
where $\sigma$ denotes the permutations.\footnote{A permutation is a special case of injective assignment. The functions $\sigma$ are all injective and associate to each number $k$ a number $j$. Therefore, in the sum, $\tL_{x_{\sigma^{-1}(j)}}$ means the $\tL_{x_k}$ such that $\sigma(k)=j$.} Finally, when $m>n$ each sequence surviving the PSL limit contains all the $\tL_{x_k}$ and $m-n$ $\hH^{(W)}$. Mathematically, the sequences appearing in the sums are the injective maps from an $n$-element set (that of the $\tL_{x_k}$ with $k=1,\dots,n$) to an $m$-element set (that of the times $s_j$ with $j=1,\dots,m$), where all empty slots are filled by $\hH^{(W)}$. The number of terms in the sum is therefore $m!/(m-n)!$. Denoting such injective assignment by $\sigma$ and by $\textrm{Im}\ \sigma$ its image in the $m$-element set, we define
\begin{equation}
\hO_j^{(\sigma)} (s_j):=
\begin{cases}
\tL_{x_{\sigma^{-1}(j)}} (s_j), 
&\quad j \in \textrm{Im}\ \sigma,
\\
\hH^{(W)} (s_j), &\quad j \notin \textrm{Im}\ \sigma,
\end{cases}
\end{equation}
so that we have
\begin{equation}
\frac{(-i)^m}{m!}\bra{1_n}\mcT_{\hW_n}
\prtg{\hW_n(s_1)\cdots\hW_n(s_m)}\ket{0_n}
\xrightarrow{\text{PSL limit}}
\frac{(-i)^m}{m!}\Upsilon^{n/2}\lambda^{(m-n)/2}
\mcT\prtg{\sum_\sigma \prod_{j=1}^{m}\hO_j^{(\sigma)} (s_j)}.
\end{equation}

All of the above can be summarized by defining
\begin{equation}
\hM_{n,m} (s_1,\dots,s_m)
:=
\lim_{\Upsilon\to 0}
\frac{\bra{1_n}\mcT_{\hW_n}
\prtg{\hW_n(s_1)\cdots\hW_n(s_m)}\ket{0_n}}{\Upsilon^{n/2}\lambda^{(m-n)/2}}
=
\begin{cases}
0, &\quad m<n,\\
\mcT\prtg{\sum_\sigma \prod_{j=1}^{m}\hO_j^{(\sigma)} (s_j)}, 
&\quad m\geq n,
\end{cases}
\end{equation}
so that
\begin{equation}
\frac{(-i)^m}{m!}\bra{1_n}\mcT_{\hW_n}
\prtg{\hW_n(s_1)\cdots\hW_n(s_m)}\ket{0_n}
\xrightarrow{\text{PSL limit}}
\frac{(-i)^m \Upsilon^{n/2}\lambda^{(m-n)/2}}{m!}\hM_{n,m} (s_1,\dots,s_m),
\end{equation}
where, as shown in Sec.~\ref{Sec:FlashProbabilities}, the factor $\Upsilon^{n/2}$, appearing quadratically in Eq.~\eqref{eq:MultiPointProbabilityFlashesPSL}, will combine with $\mu_C^n$ to give $\lambda^n$. It is interesting to mention that $\hM_{n,m}$ contains a sum of $m!/(m-n)!$ terms and is also divided by $m!$. Therefore, it remains a factor $1/(m-n)!$ which helps the convergence of the series  used to compute the flash intensity of order $n$.

\end{document}